\documentclass[preprint,aps,ams, eqsecnum]{revtex4}

\begin{document}

\title{Does the Third Law of Thermodynamics hold in the Quantum Regime?}

\author{R. F. O'Connell}

\affiliation{Department of Physics and Astronomy, Louisiana State
University,
Baton Rouge, LA  70803-4001, USA}

\date{\today}

\begin{abstract} The first in a long series of papers by John T. Lewis,
in collaboration with G. W. Ford and the present author, considered the
problem of the most general coupling of a quantum particle to a linear
passive heat bath, in the course of which they derived an exact formula
for the free energy of an oscillator coupled to a heat bath in thermal
equilibrium at temperature $T$.  This formula, and its later extension
to three dimensions to incorporate a magnetic field, has proved to be
invaluable in analyzing problems in quantum thermodynamics.  Here, we address
the question raised in our title viz.  Nernst's third law of thermodynamics.
\end{abstract}

\maketitle

\section{INTRODUCTION}

In recent years, the realization that quantum effects can play an
important role in thermodynamic theory has led to an intense interest
in quantum and mesoscopic thermodynamics \cite{capek,front,quantum,nieu,ford1},
to the extent that some authors \cite{capek,front,quantum,nieu} have
questioned whether the fundamental laws of thermodynamics remain valid.  The
arguments are often very subtle and sometimes not easily dismissed. 
Here, our goal is not to survey the whole area but, instead, to focus in
depth on questions which have been raised in relation to Nernst's third
law of thermodyanmics, which states that the entropy $S$ vanishes as the
temperature
$T\rightarrow{0}$.  Ford and the present author \cite{ford1} already
considered this question for the one-dimensional problem of a quantum
oscillator in an arbitrary heat bath and also briefly commented on what
we considered to be the flaw in the calculations of other authors.  We
now expand on these remarks in Sec. II (where we discuss various
approaches to the calculation of $S$), by carrying out a detailed
calculation in which we demonstrate explicitly how use of approaches
based on an incorrect utilization of the Wigner function formalism, and also
the von Neumann formula, leads to an incorrect form of  a well-known result. 
By contrast, the method we employed in \cite{ford1} was based on calculating
$S$ from an exact expression for the free energy
$F$ calculated by us in 1985 at the start of a long collaboration with
John T. Lewis \cite{ford2}.  This formula for $F$ was later extended to
three dimensions to incorporate a magnetic field \cite{li}.  Thus, in Sec.
III, we review these results and explicitly write down the results for
$F$ and
$S$ in the three-dimensional case.  All of these results involve just a
single integral and, in Sec. IV, we evaluate this integral for the case
of an Ohmic bath and low temperature.  This enables us to confirm
Nernst's law.  Our conclusions are given in Sec. V.

\section{Calculation of the Entropy $S$}

For a quantum particle in a heat bath, a basic quantity is the density
matrix
\begin{equation}
\rho =e^{-H/kT}/\mathrm{Tr}\{e^{-H/kT}\},
\end{equation}  where \textit{H} is the Hamiltonian for the whole system
(quantum particle plus heat bath plus interaction).  The question is how to
calculate the entropy $S$ from this expression.  In Ref. 4, three different
results for the entropy are obtained based on use of the free energy ($S_{p}$
given in (3.59) of \cite{nieu}), the Wigner distribution ($S_{B}$ given in
(4.5) and (4.38) of \cite{nieu} and referred to as the "Boltzmann" energy)
and the von Neumann expression ($S_{vN}$ given in (4.36) of \cite{nieu}).  In
particular, the authors of \cite{nieu} argue that " - - neither the von
Neumann entropy nor the Boltzmann entropy vanishes when the bath temperature
is zero", leading them to note " - - the violation of the third law reported
here for nonweak coupling".  On the contrary, we argue that all such
calculations should lead to the same result so we are motivated to examine
the question of what went wrong.  In the process, we will argue that a
calculation based on the free energy is the simplest way to proceed to obtain
the correct result, particularly if one uses the formula given in
\cite{ford2}.

First, we examine the calculation of $S_{vN}$, which was based on applying
the von Neumann formula to the \underline{reduced} density matrix.  But as we
pointed out in \cite{ford1}, for a system with a non-negligible interaction
energy, " - - the von Neumann formula can only be applied to the
\underline{entire} system and not to the reduced system".  At first glance,
these may seem to be at variance with the fact that numerous calculations rely
on the calculation of the von Neumann entropy of mixed states obtained by
tracing out other variables.  The answer is that this procedure is correct
only when the interaction energy is negligible, as is often the case (so that
the total entropy is equal to the sum of the entropies of the parts). 
However, this is not so in the case of heat baths (where we are dealing with
\underline{non-additivity of entropies}), as may be verified explicitly. 
This is what motivated our calculation with John Lewis, where the free
energy $F$ of the electron in the presence of the heat bath was calculated
by first using the von Neumann formula to calculate the free energy of the
\underline{entire interacting} system and then subtracting the well-known
free energy of the heat bath itself.  Simply differentiating - $F$ with
respect to $T$ [see [2.10) below] leads to the correct result $S_{p}$.  The
authors of \cite{nieu} used our procedure in their calculation of $S_{p}$
[their Eq. (3.59)], obtaining a correct result in agreement with our previous
calculation of this quantity \cite{ford1}, but in disagreement with their
result [given in their Eq. (4.36)] for $S_{vN}$ (which was obtained assuming
additivity of entropies).  If, instead, one uses the density matrix of the
whole system, the corresponding result will coincide with the result
$S_{p}$.

Second, we examine the calculation of $S_{B}$ of \cite{nieu} which was based
on writing down the von Neumann formula for the entropy,
$S=-k\mathrm{Tr}\{\rho \log \rho \}$, and then making use of the
\underline{reduced} Wigner distribution function $W$ \cite{hill} corresponding
to $\rho $, by simply replacing $\rho $ with $W$ in the von Neumann
expression, as may be verified by examination of equations (4.3) and (4.4)
of \cite{nieu}, leading to the authors' equation (4.5).  The calculation
within this framework is correct but, unfortunately, the framework is not
correct, as we shall now demonstrate explicitly in the simplest case where the
interaction energy is taken to be negligible (i.e. by letting the dissipation
parameter $\gamma\rightarrow{0}$), where well-known results for the entropy of
a \underline{single} oscillator in equilibrium at temperature $T$ \cite{fey}
may be used as a benchmark.  For an oscillator at temperature $T$, we
have the well-known exact result for the Wigner distribution \cite{hill}
\begin{equation}
W(q,p)=N\exp(-aH), \label{ear123}
\end{equation} where
\begin{equation}
\frac{1}{a}=\frac{\hbar\omega}{2}\coth\left(\frac{\hbar\omega\beta}{2}\right),
\label{ear124}
\end{equation}
\begin{equation} N=\frac{\omega
a}{2\pi}=\left(\pi\hbar\right)^{-1}\tanh\left(\frac{\hbar\omega\beta}{2}\right)\equiv
\left(2\pi\hbar\right)^{-1}N_{1}, \label{ear125}
\end{equation} and
\begin{equation}
H(q,p)=\frac{p^{2}}{2m}+\frac{1}{2}m\omega^{2}q^{2}. \label{ear126}  
\end{equation} Also, we note that
\begin{equation}
\int\int dqdp~W(q,p)=1,
\end{equation} and 
\begin{equation}
<H(q,p)>=\int\int dqdp~W(q,p)~H(q,p)=\frac{1}{a}.
\end{equation}
If we now replace (incorrectly as we shall see) $\rho$
in the von Neumann formula by $2\pi\hbar$ ${W}$ then we obtain for the
entropy (which we denote here as $S^{(W)}$, indicating that it was
derived using the Wigner distribution)
\begin{eqnarray} (S^{(W)}/k) &=& -\int\int dqdp~W(q,p)\log\left[2\pi\hbar
W(q,p)\right] \nonumber \\  &=& -\log
N_{1}\int\int dqdp~W(q,p)+a\int\int dqdp~H(q,p)W(q,p) \nonumber \\  &=&
-\log N_{1}-aN\frac{\partial}{\partial a}\int\int dqdp\exp(-aH) \nonumber
\\  &=& \log\left[\frac{1}{2}\coth
\frac{\hbar\omega\beta}{2}\right]+1. \nonumber\\
\end{eqnarray} This result is incorrect, being in striking contrast
with the correct result given by \cite{fey} 
\begin{equation} 
(S/k)=-\log\left[2\sinh (\hbar\omega
/2kT)\right]+\frac{\hbar\omega}{2kT}\coth (\hbar\omega/2kT).
\end{equation}  Only in the high temperature classical regime do equations
(2.8) and (2.9) agree, in which case the Wigner distribution function
simply reduces to a classical distribution function.  One of the problems with
this method stems from the fact that the Wigner distribution corresponding to
$log~\rho$ is not the same as the $log$ of the Wigner distribution
corresponding to
$\rho$.  Thus, we need an alternative approach.  

In fact, as already noted
\cite{ford1,ford2,li}, the best approach is simply to calculate $F$
from the formula
\begin{equation}
S(T)=-\frac{\partial{F}}{\partial{T}},
\end{equation} and in the next section we discuss the method by which
$F$ is obtained.  As we shall see, the result $S_{p}$ of \cite{nieu} is
correct and agrees with a result previously obtained by us \cite{ford1} in
the case of an Ohmic heat bath in the absence of a magnetic field.  However,
we will also note that the method used in
\cite{nieu}, for the calculation of $F$, is unnecessarily complicated and only
applies to the Ohmic model whereas, by contrast, our result is expressed as a
simple integral in terms of the generalized susceptibility and applies, not
only for the Ohmic bath, but for an arbitrary heat bath and in the presence of
a magnetic field.

\section{Fundamentals}

The Hamiltonian for a charged quantum oscillator (with a force constant
$K=m\omega^{2}_{0}$) moving in an external magnetic field and linearly
coupled to a passive heat bath (consisting of an infinite number of
oscillators) may be written as
\cite{li}
\begin{eqnarray}
H_{o}~=~{1 \over{2}m}\Bigl[{\vec p}-{e \over c}{\vec
A}\Bigr]^{2}~+~{1\over{2}}K{\vec r}^{~2}~+~\sum_{j}^{ } 
\Bigl[{{\vec p}_{j}^{~2}\over{2}m_{j}}~+~{1\over{2}}m_{j} 
\omega^{2}_{j}\Bigl({\vec q}_{j}-{\vec r}\Bigr)^{2}\Bigr]. 
\end{eqnarray}  This is the independent oscillator (IO) model in the
presence of an external magnetic field ${\vec B},$ where $e, m, {\vec
p}, {\vec r}$ are the charge, mass, momentum and position of the
oscillator respectively and the corresponding quantities with the lower
indices $j$ refer to the $j$th heat bath oscillator.~~ The vector
potential ${\vec A}$ is related to the magnetic field ${\vec B}$ through
the equation
\begin{equation}
{\vec B} ({\vec r}) ~=~ {\vec \nabla} \times {\vec A} ({\vec r})~.
\end{equation}
Next, using the Heisenberg equations of motion, the problem has been
formulated exactly in terms of the quantum Langevin equation.
\begin{equation}
m{\ddot{\vec{ r}}} ~+~ \int^{t}_{- \infty} dt^{\prime} \mu (t - t^{\prime})
{\dot{\vec{r}}}(t^{\prime}) ~-~ {e \over c} {\dot{\vec{ r}}} \times {\vec B}
~+~ K{\vec r} ~=~ {\vec F}(t). 
\end{equation}  Here $F(t)$ is the fluctuation force (whose exact form is
known but is not relevant to our present discussion) and $\mu(t)$ is the
so-called memory (non-Markovian) term given by \cite{li}
\begin{equation}
\mu(t)=\sum_{j}m_{j}\omega^{2}_{j}~\cos(\omega_{j}t)\theta(t),
\end{equation} where $\theta(t)$ is the Heaviside function.  

We note that
$\mu(t)$ does not depend on the magnetic field.  Fourier transforming
(3.3), we obtain \cite{li}
\begin{equation}
{\tilde r}_{\rho}(\omega ) ~=~  \alpha_{\rho \sigma} (\omega) ~[{\tilde
F}_{\sigma} (\omega)]  ~,
\end{equation} where 
\begin{equation}
\alpha_{\rho \sigma} (\omega) \equiv [ D (\omega)^{-1}]_{\rho
\sigma} ~=~ \Bigl[\lambda^{2} \delta_{\rho \sigma} - \Bigl( \omega {e \over
c} \Bigr)^{2} ~B_{\rho} B_{\sigma} ~-~ \epsilon_{\rho \sigma \eta}
B_{\eta}  \lambda i \omega {e \over c} \Bigr]/det  D (\omega)  ~,
\end{equation} is the generalized susceptibility, with 
\begin{equation}
det  D(\omega) ~=~ \lambda \Bigl[ \lambda^{2} - (\omega {e \over
c})^{2} ~{\vec B}^{2} \Bigr]  ~, 
\end{equation} and 
\begin{equation}
\lambda  (\omega ) ~=~ - m\omega^{2} ~+~ K - i \omega {\tilde \mu}
(\omega)  ~ \equiv ~ \{ \alpha^{(0)} (\omega ) \}^{-1}  ~.
\end{equation} Also 
\begin{equation}
{\tilde \mu} (\omega) \equiv \int^{\infty}_{o} dt  e^{i \omega t} \mu
(t), 
\end{equation}
\begin{equation}
{\tilde r}_{\sigma}(\omega) ~=~ \int^{\infty}_{- \infty} dt e^{i \omega
t} r_{\sigma}(t),
\end{equation} and so on, ~  and where ~ $\epsilon_{\rho \sigma \eta}$ ~ is
the Levi-Civita symbol, a totally antisymmetric tensor.  ~~ 
The Greek indices stand for three spatial
directions (i.e. $\rho$, $\sigma$, etc. ~=~ 1,2,3) and we adopt Einstein
summation convention for repeated Greek indices.  From (3.4) and (3.9), we
obtain the Fourier transform of the memory function  
\begin{equation}
{\tilde \mu} (\omega) ~=~ {i \over 2} \sum_{j}^{ } m_{j} \omega_{j}^{2}
\Bigl[ {1 \over \omega - \omega_{j} } ~+~ {1 \over \omega + \omega_{j}}
\Bigr]. 
\end{equation} We have now all the tools required in order to calculate
various quantities.  Here, we discuss the free energy and associated entropy.

The free energy ascribed to the oscillator,
$F(T)$, is given by the free energy of the system minus the free energy of
the heat bath in the absence of the oscillator.  This is a non-trivial
quantity to calculate, details of which may be found in \cite{li} leading to
the result
\begin{equation}
F_{o}(T,B) ~=~ {1 \over \pi} \int^{\infty}_{o} d\omega f(\omega ,T) Im
\Bigl\{ {d \over d \omega } ln \Bigl[ det \alpha (\omega + io^{+}) \Bigr]
\Bigr\},
\end{equation} where $f(\omega ,T)$ is the free energy of a single
oscillator of frequency $\omega $, given by
\begin{equation} 
f(\omega ,T)=kT\log[1-\exp\left(-\hbar\omega/kT\right)]. 
\end{equation} Here the zero-point contribution $(\hbar \omega /2)$ has
been omitted and
\begin{equation}
{\rm det} \alpha (\omega ) ~=~ [\alpha^{(0)} (\omega )
]^{3} \left[1 - \Bigl[ {eB\omega \over c}  \Bigr]^{2} [\alpha^{(0)}
(\omega ) ]^{2} \right]^{-1},
\end{equation} with
\begin{equation}
 [ \alpha^{(0)} (\omega ) ]^{-1} ~=~ - m\omega^{2} ~+~ K - i \omega
{\tilde \mu} (\omega).
\end{equation}  This enabled us to write
\begin{equation}
F_{o}(T,B) ~=~ F_{o} (T,0) ~+~\Delta F_{o} (T,B),   
\end{equation} where
\begin{equation}
F_{o} (T,0) ~=~ {3 \over \pi }
\int^{\infty}_{o} ~ d \omega ~  f(\omega ,T) Im \Big\{ {d \over d \omega}
ln \alpha^{(0)} (\omega )  \Bigr\}
\end{equation} is the free energy of the oscillator in the absence of the
magnetic  field (in agreement with Eq. (5) of Ref. 6, except for the
extra factor of 3 which results from our consideration here of three
dimensions) and the correction due to the magnetic field is given by
\begin{equation}
\Delta F_{o}(T,B) ~=~-{1\over\pi}~\int^{\infty}_{o}~d \omega~ 
f(\omega , T) Im\Big\{{d\over{d}\omega} ln  [1-\bigl({eB \omega
\over{c}} \bigr)^{2} (\alpha^{(0)}(\omega))^{2}] \Big\} ~.
\end{equation}

\section{Ohmic Heat Bath}

In the case of the Ohmic heat bath, ${\tilde\mu}(\omega)~=~m
\gamma,$ ~a constant, which is the simplest memory function one can
choose and is an oft-used \cite{nieu} choice.  Thus, making use of (3.7) and
(3.8),~(3.3) becomes
\begin{eqnarray} 
F_{o} (T,B) &=& {1\over\pi}~\int^{\infty}_{o}d\omega{f}(\omega , T)
\\ & &
 \Bigg\{ {\gamma (\omega^{2} + \omega^{2}_{o})
\over (\omega^{2} - \omega^{2}_{o})^{2} + \omega^{2} \gamma^{2} } +
{\gamma (\omega^{2} + \omega^{2}_{o}) \over (\omega^{2}
-\omega^{2}_{o} + \omega_{c} \omega)^{2} + \omega^{2} \gamma^{2}}
+{\gamma (\omega^{2} + \omega^{2}_{o}) \over (\omega^{2}
- \omega^{2}_{o} - \omega_{c} \omega )^{2} +\omega^{2} \gamma^{2}}
 \Bigg\} \nonumber,
\end{eqnarray} where $\omega_{o} ~=~ (K/m)^{1 \over 2}$ and
$\omega_{c}=(eB/mc).$  

We now wish to examine the low-temperature behavior of this result. 
First, we note that the function $f(\omega,T)$ vanishes exponentially for
$\omega\gg{k}T/\hbar$. Therefore as $T\rightarrow{0}$ the integrand is
confined to low frequencies and we can obtain an explicit result by
expanding the factor multiplying $ f(\omega ,T)$ in powers of $\omega$. 
Hence, in particular, we see that the terms involving the magnetic field
(the $\omega_{c}$ terms) are negligible since they always contain an
$\omega$ factor.  Thus our result corresponds to the results obtained in
the absence of a magnetic field \cite{ford2}, except for the extra factor
of 3 which results from our consideration here of three dimensions.  Thus,
proceeding as in \cite{ford1}, we obtain in the low-temperature case
\begin{eqnarray}
F(T) &\cong& \frac{3\gamma kT}{\pi \omega
_{0}^{2}}\int_{0}^{\infty }d\omega
\log \left[ 1-\exp \left( -\hbar \omega /kT\right) \right] \nonumber \\
&=& -\frac{\pi }{2}\hbar \gamma \left( \frac{kT}{\hbar\omega
_{0}}\right)^{2},
\end{eqnarray} so that
\begin{equation}
S(T)=-\frac{\partial F}{\partial T}=\pi\gamma\frac{k^{2}T}{\hbar\omega
_{0}^{2}}.
\end{equation}  Hence $S(T)\rightarrow 0$ as $T\rightarrow 0$, in conformity
with the third law of thermodynamics.  We also note that the result for the
entropy is independent of the magnetic field $B$ and actually corresponds
(except for a factor of 3 because here we considered 3 dimensions) to the
result obtained earlier by us \cite{ford1} in the $B=0$ case and also agrees
with the result for $S_{p}$ given in \cite{nieu}.  The latter calculation was
also based on a determination of the free energy $F$ but this calculation was
much more complicated, involving a non-trivial determination of the
frequencies of the interacting system \cite{ford3}.  By contrast, our
calculation was based on a simple integral \cite{ford2,li}, given here by
(3.12), and which involves only the specification of the generalized
susceptibility
$\alpha(\omega)$.

\section{Conclusion}

Similar results to those obtained in the Ohmic case may be obtained also
in the case of a blackbody radiation heat bath and, in fact, in
the case of arbitrary heat baths.  Thus, we conclude that for the case of
a quantum oscillator coupled to an arbitrary heat bath, we have shown that
Nernst's third law of thermodynamics is still valid: the entropy vanishes
at zero temperature. In this connection we should emphasize that the
basis of our discussion is the remarkably simple formula for the free energy
of an oscillator, in an arbitrary heat bath at arbitrary temperature $T$,
obtained in the 1985 paper \cite{ford2} with J. T. Lewis.  Since the
validity of the second law of thermodynamics has also been called into
question \cite{capek,front,quantum,nieu}, we note that this formula for the
free energy was also used in refuting speculations that quantum effects could
lead to extraction of energy from a zero-temperature heat bath \cite{ford4}.

\section{Acknowledgment}

The author is pleased to acknowlege a long-time collaboration with G. W. Ford
which led to the essential results presented here.

\section{Dedication}

The author dedicates this paper to the memory of John T. Lewis, a
collaborator over two decades and a wonderful friend.  He is sorely
missed.

\end{document}